\documentclass[aip,pop,onecolumn,preprint,superscriptaddress,amsmath,amssymb,showpacs]{revtex4}

\usepackage{graphicx}

\begin{document}


\title{High-quality proton bunch from laser interaction with a gas-filled cone target} 


\author{H.Y.Wang} \affiliation{State Key Laboratory of Nuclear
Physics and Technology, Peking University, Beijing 100871, China}
\affiliation{Key Lab of High Energy Density Physics Simulation,
CAPT, Peking University, Beijing 100871, China}

\author{F.L.Zheng}
\affiliation{Key Lab of High Energy Density Physics Simulation,
CAPT, Peking University, Beijing 100871, China}
\author{Y.R.Lu}
\affiliation{State Key Laboratory of Nuclear Physics and Technology,
Peking University, Beijing 100871, China}
\author{Z.Y.Guo}
\affiliation{State Key Laboratory of Nuclear Physics and Technology,
Peking University, Beijing 100871, China}
\author{X.T.He}
\affiliation{Key Lab of High Energy Density Physics Simulation,
CAPT, Peking University, Beijing 100871, China}

\author{J.E.Chen} \affiliation{State Key Laboratory of
Nuclear Physics and Technology, Peking University, Beijing 100871,
China}
\author{X.\,Q. Yan} \email[]{x.yan@pku.edu.cn}
\affiliation{State Key Laboratory of Nuclear Physics and Technology,
Peking University, Beijing 100871, China}\affiliation{Key Lab of
High Energy Density Physics Simulation, CAPT, Peking University,
Beijing 100871, China}


\date{\today}

\begin{abstract}
Generation of high-energy proton bunch from interaction of an
intense short circularly polarized(CP) laser pulse with a gas-filled
cone target(GCT) is investigated using two-dimensional
particle-in-cell simulation. The GCT target consists of a hollow
cone filled with near-critical gas-plasma and a thin foil attached
to the tip of the cone. It is observed that as the laser pulse
propagates in the gas-plasma, the nonlinear focusing will result in
an enhancement of the laser pulse intensity. It is shown that a
large number of energetic electrons are generated from the
gas-plasma and accelerated by the self-focused laser pulse. The energetic
electrons then transports through the foil, forming a backside
sheath field which is stronger than that produced by a simple planar
target. A quasi-monoenergetic proton beam with maximum energy of 181
MeV is produced from this GCT target irradiated by a CP laser pulse
at an intensity of $2.6\times10^{20}W/cm^2$, which is nearly three
times higher compared to simple planar target(67MeV).

\end{abstract}

\pacs{52.38.Kd, 41.75.Jv, 52.35.Mw, 52.59.-f}

\maketitle 

\section{Introduction}
With the rapid development of the chirped pulse amplification
technique, generation of energetic ion beam by interactions of an
ultra intense laser pulse with a solid target has become realizable.
Such energetic ions can be promising for many scientific or societal
applications, such as proton radiography\cite{Borghesi2002}, fast
ignition for inertial confined
fusion\cite{Naumova2009,Roth2001,Temporal2002}, or
hadron-therapy\cite{Khoroshkov2002}. For most of these applications,
ion beams with high energy, low energy spread and high collimation
are required.

Depending on the target paraments and laser intensity, ions can be
accelerated by several different mechanisms, such as shock
acceleration\cite{Denavit1992,Silva2004},light-pressure
acceleration\cite{Yan2008,Yan2009b, Qiao2009,Henig2009}, Coulomb
explosion\cite{Fourkal2005}, target-normal sheath acceleration
(TNSA)\cite{Wilks2001,Snavely2000,Fuchs2006}, etc., as well as their
combinations. In TNSA, the energetic electrons produced at the front of a
target by the laser ponderomotive force propagate through the target
into the backside vacuum can generate a sheath electrostatic field.
The sheath field, of order $10^{12}$V/m, can accelerate the ions on
the target back surface to high energies. However, the proton beams
obtained in this way are typically characterized by low particle
density, large divergence, and almost $100\%$ energy spread. An
improved TNSA scheme, using a microstructured double-layer (DL)
target, can  decrease the energy spread. The possibility to generate
1.3 MeV proton beams with energy dispersion$\thicksim25\%$ and 3 MeV
carbon beams with energy dispersion $\thicksim17\%$ using a
microstructured DL target has already been demonstrated
experimentally by Schwoerer et al.\cite{Schwoerer2006} and Hegelich
et al.\cite{Hegelich2006}, respectively.

A tiny hollow metal cone was first introduced in fast ignition
experiments to shield the igniting laser pulse from the underdense
region of the precompressed fuel plasma\cite{Kodama2001}, and a
remarkable increase in the thermal fusion-neutron yield was
observed. Since then the cone target was intensively examined both
in experiments and
simulations\cite{Chen2005,Stephens2003,Woerkom2008,Mason2006,Lei2006,Pasley2007,Sakagami2006,Nagatomo2007,Key2007,Cai2009,King2009,Green2007,Rassuchine2009}.
PIC simulations showed that a cone target could nonlinearly guide
and focus a laser beam, and improve the efficiency of the coupling
and transport of the energy into dense
plasma\cite{Sentoku2004,Nakamura2007}. Accelerating proton beams
using a cone target with open tip was also studied by Cao et
al.\cite{Cao2008}, and energetic ion bunches of high density were
observed.
 \begin{figure}[!ht]{
 \includegraphics[width=0.9\columnwidth]{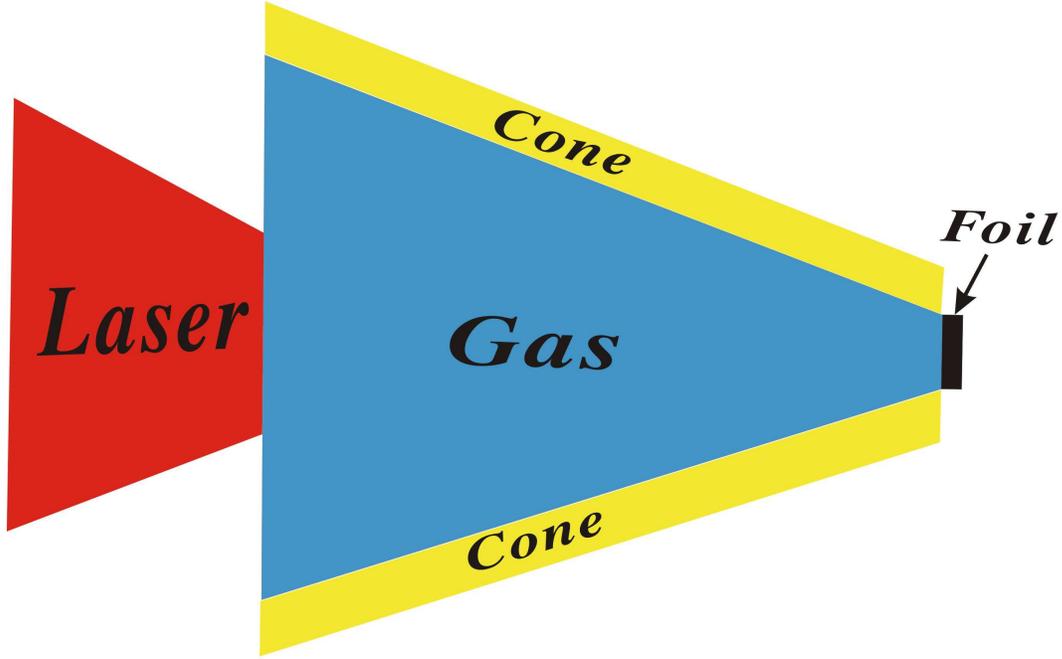}
  \caption{(color online).Schematic view of the interaction of a laser pulse with a
GCT target. The left trapezoid represents the laser pulse (red), the
trapezoid irradiated by the laser pulse represents the gas-plasma
(blue), the two parallelograms besides the gas-plasma represent the
cone (orange), the rectangle on the tip of the cone represents the
foil (dark). } \label{fig1}}
\end{figure}
In this paper, we report that quasi-monoenergetic proton beam with
peak energy of 130MeV and maximum energy of 181 MeV can be generated
from a gas-filled cone target irradiated by a CP Gaussian laser
pulse at an intensity of $2.6\times 10^{20}W/cm^2$. The gas-filled
cone target, as shown in Fig.\ref{fig1}, consists of a hollow cone
filled with near-critical heavy-ion gas-plasma and a thin foil
attached to the tip of the cone. Our results indicate that,
comparing with that from a simple proton target, energetic protons
with smaller energy spread and higher energy can be obtained. This
result can be attributed to the much higher electron density and
temperature behind the foil and the small transverse size of the
foil. The energetic electrons are generated from the gas-plasma and
accelerated by the enhanced laser pulse, which undergoes
self-focusing in gas-plasma and is even focused by the tip of the
cone. These energetic electrons can easily propagate through the
thin foil to form a stronger sheath field behind the foil (than that
behind a planar target). Since the foil has a small transverse size,
the protons in the foil are accelerated in the homogenous sheath
field, so that the protons are accelerated longitudinally forward
with smaller energy spread. In contrary to TNSA acceleration, here
energetic electrons originate mainly from the gas rather than the
solid target.


\section{Simulation Parameters}
We carried out simulation using a fully relativistic
particle-in-cell code (KLAP2D) \cite{zheng2002,Yan2008}.In
simulations, the simulation box is $80\lambda\times20\lambda$, where
$\lambda=1\mu m$ is the laser wavelength, and contains
$3200\times800$ cells. A CP
 laser pulse with a peak laser intensity of
$I_L=2.6 \times10^{20} W/cm^{2}$ is normally incident from the left
side, The pulse has a Gaussian radial profile with
$2\sigma=10\lambda$ full width at half maximum and a trapezoidal
shape longitudinally with $40T$ flat top and $1T$ ramps on both
sides, where T is the laser period.
  The corresponding peak dimensionless laser amplitude $a_0=eE/(m_e \omega
c)$ is 9.8, where E, $\omega$, c,$m_e$, and e are the laser electric
field, frequency, speed of light in the vacuum, electron mass, and
charge, respectively.  The GCT target, as shown in Fig.\ref{fig1},
consists of electrons, protons, and heavy carbon ions.
\begin{figure}[!ht]
  \includegraphics[width=1.0\columnwidth]{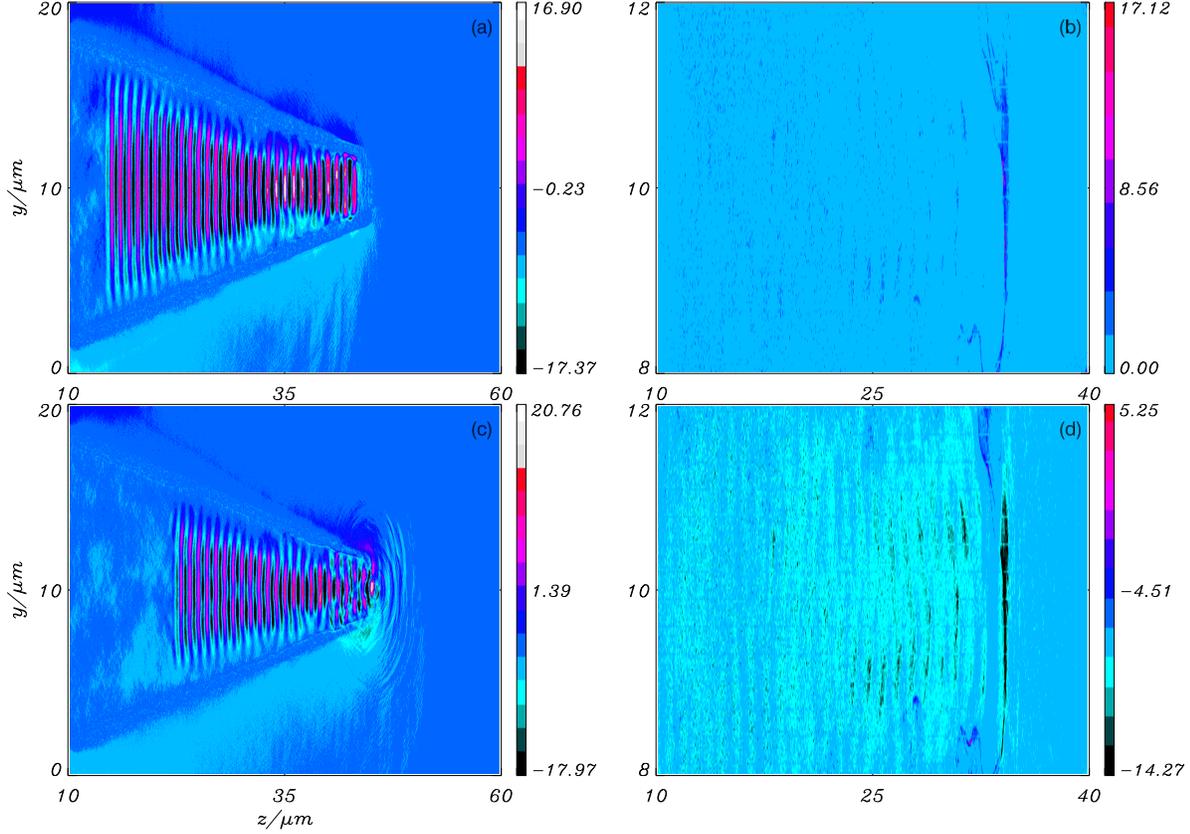}
  \caption{ (color online). Left: Transverse electric field (in units of $m_ec\omega/e$) at (a)$t=56T$, (c) $t=64T$. Right: (b) electron density (in units of critical density $n_c$), and (d) the longitudinal electron current ${J_z}_e$ (in units of $en_cc$) at $t=44T$.}
 \label{fig2}
\end{figure}
The initial temperature of electrons, protons, and carbon ions is 10
eV. The cone has a width of $1\mu m$, and is located in $10< z[\mu
m]<45$ with the diameters of the left and right cone openings of
$16\mu m$ and $2\mu m$, respectively. For simplicity, the cone
consists of carbon plasma with an electron density $n_e=10n_c$,
where $n_c=\pi m_ec^2/(e\lambda)^2$ is the critical density. The
carbon gas-plasma is full in the cone with density $n_e=0.8n_c$. The
foil with $2\mu m$ wide and $0.35\mu m$ thick is placed at $z=45\mu
m$.  It consists of a proton-carbon mixed plasma with an electron
density $n_e=40n_c$,  and the ratio of C:H=1:1.

\section{Simulation Results}

A laser beam propagating in underdense plasma with a frequency
$\omega_p$ smaller than the laser frequency $\omega$ undergoes
relativistic
self-focusing\cite{Mori1988,Pukhov1996,Mori1997,Asthana2000,Tripathi2010}
as soon as its total power P exceeds the critical value
\begin{equation}
 P_{cr}\approx17(\omega/\omega_p)^2GW;
 \label{equ1}
\end{equation}

The self-focusing is due to the relativistic mass increase of plasma
electrons and the ponderomotive expulsion of electrons from the
pulse region. Both effects lead to a local decrease of plasma
frequency and an increase in refractive index. The strong non-linear
self-focusing of the laser pulse propagating in the near-critical
gas-plasma at $t=56T$ is shown in Fig.\ref{fig2}(a). For clarity,
only a part of the simulation box is shown. The spot size of the
laser is focused to be smallest at $z=35\mu m$, and the transverse
electric field is enhanced to 17 there, which is 1.7 times higher
than the initial laser electric field. The pulse retains its
Gaussian radial profile, however, its spot size varies with the
distance of propagation in a periodic manner. The smallest spot size
at $z=35\mu m$ is about $3\mu m$, while it varies to about $4\mu m$
at $z=40\mu m$. This result is due to dynamic balance between
diffraction and non-linear self-focusing, which is also in good
agreement with the analysis of the paraxial ray
approximation\cite{PANWAR2009}.  When the laser propagates to the
tip of the cone, it is even focused or squeezed by the tip of the
cone for the small radius there, as is shown in Fig.\ref{fig2}(b).
The spot size is focused to about $1\mu m$ at the tip of the
cone($z=45\mu m$), with the transverse electric field as high as 20
there.
\begin{figure}[!ht]
  \includegraphics[width=1.0\columnwidth]{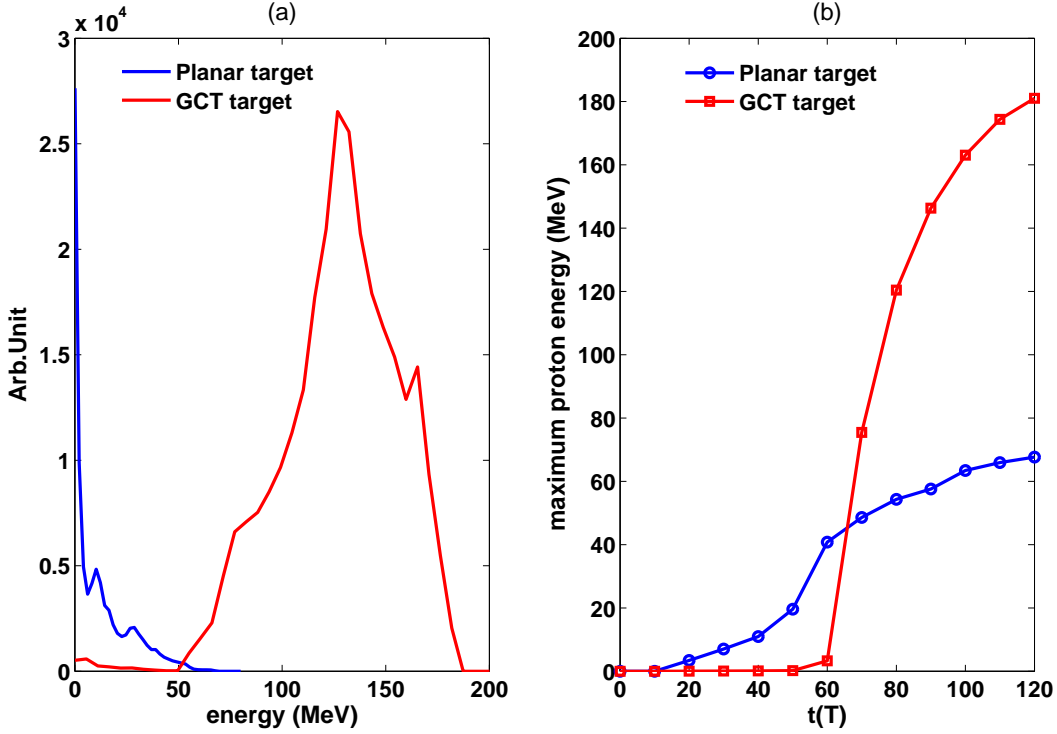}
\caption{(color online).(a) the proton energy spectrum of the
protons  behind the targets for different cases at t=120T.and (b)
Evolution of the maximum proton energy for different cases.
}\label{fig3}
\end{figure}
The electrons that are initially at the front of the pulse are more
efficiently accelerated as the pulse undergoes intensity enhancement
due to self-focusing. Strong flows of relativistic electrons,
axially comoving with the laser pulse, are observed in the
simulation, as shown in Fig.\ref{fig2}(b)and Fig.\ref{fig2}(d). The
maximum electron density near the axis is as high as $17n_c$, and
the longitudinal electron current ${J_z}_e=-en_e{v_z}_e$ is about
$-14en_cc$ (negative ${J_z}_e$ due to negative electron charge).
These energetic electrons then transport
 through the thin foil and form a strong backside sheath field there.

Fig.\ref{fig3}(a) shows the energy spectrum of the proton bunches
behind the targets in the two cases at $t=120T$. For the case shown,
the maximum energy for the GCT target is about 181 MeV, which is
nearly three times higher than that of the planar target(65MeV)under
the same conditions. The energy conversion efficiencies from laser
to protons are $2.5\%$ and $0.7\%$ for GCT and planar targets at
$t=120T$, respectively. For the GCT target, due to the small
transverse size of the foil where the sheath field is homogenous,
the energy spectrum of the proton bunch has a quasi-monoenergetic
peak with  energy dispersion of about $36\%$.
\begin{figure}[!ht]
  \includegraphics[width=1.0\columnwidth]{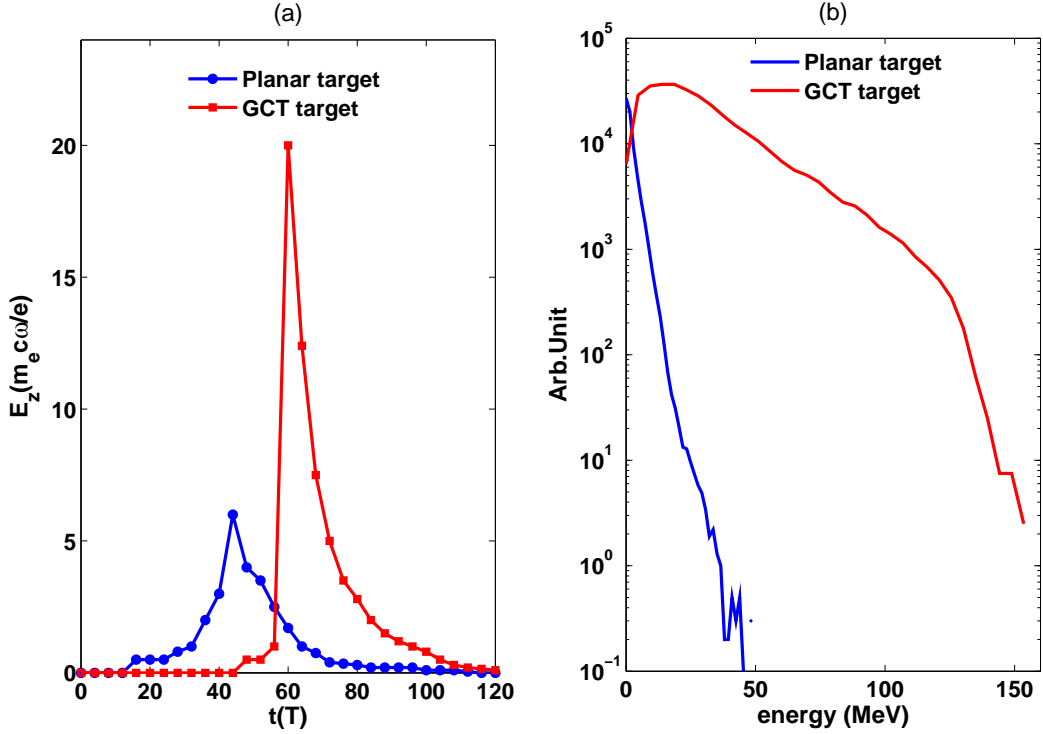}
\caption{(color online).(a)Evolution of the electrostatic fields at
the place of the proton layer for the two cases of planar and GCT
target and (b)the electron energy spectrum of the protons  behind
the targets at t=50T for planar target and t=80T for GCT
target.}\label{fig4}
\end{figure}
In contrast, the energy spectrum from the planar target is much
broadened due to multidimensional effects such as hole boring and
other instabilities. The evolution of the maximum proton energy is
shown in Fig.\ref{fig3}(b). For the planar target, as the laser
impinges on the target at $t=10T$(the planar target is initially
located at $z=10\mu m$), the maximum proton energy increases earlier
than the GCT target(for which the laser impinges on the foil at
$t=60T$). However, as the electrostatic field at the place of the
proton layer is much weaker for the planar target(see in
Fig.\ref{fig4}(a)), the increase of proton energy is much slower
than the GCT target. For the GCT target, the maximum proton energy
increases rapidly from 3.3MeV to 120MeV in only 20T(from $t=60T$ to
$t=80T$), which is attributed to the strong electrostatic field
during that time. At later time, the maximum proton energies in both
cases remain almost constant.

\begin{figure}[!ht]
  \includegraphics[width=1.0\columnwidth]{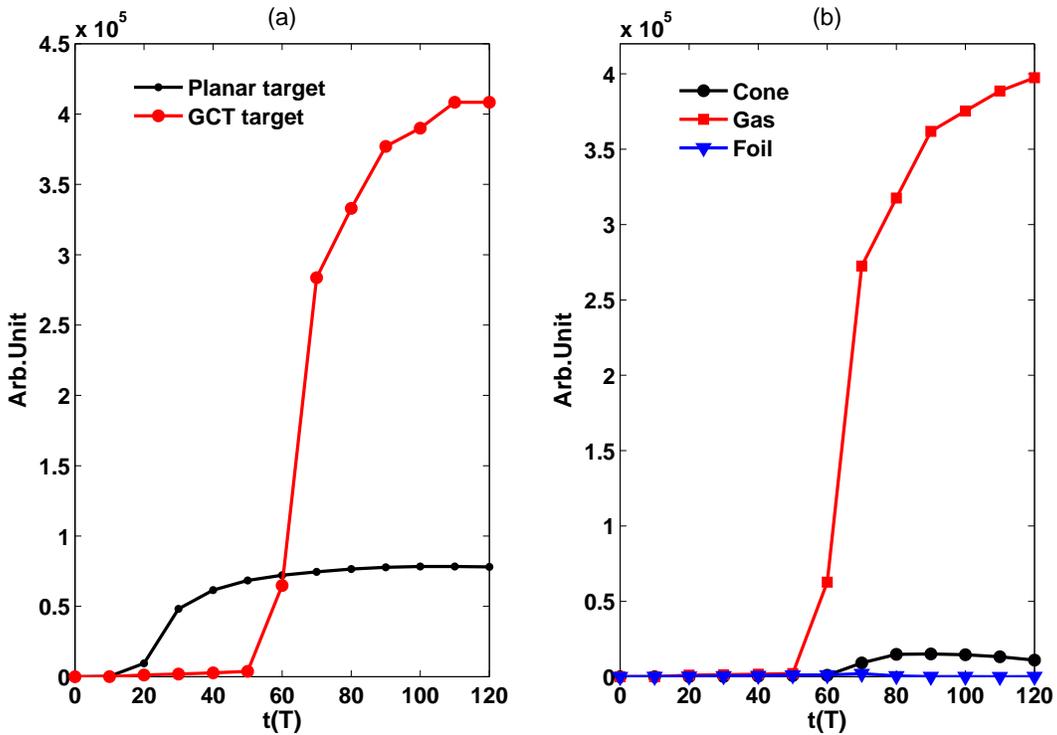}
\caption{(color online).(a) Time evolution of the number of
electrons behind the planar and GCT targets. (b)Time evolution of
the number of electrons behind the planar originated from different
places. }\label{fig5}
\end{figure}

Evolutions of the electrostatic fields at the place of the proton
layer for the planar and GCT targets are shown in Fig.\ref{fig4}(a),
which explains the energy enhancement of GCT target in
Fig.\ref{fig3}(b). The electrostatic fields straight up quickly at
$t=60T$ for the GCT target when the laser impinges on the target,
because the energetic electrons generated from the gas-plasma reach
the back side of the foil at $t=60T$ and establish a strong sheath
field there(see in Fig.\ref{fig5}(a)). The maximum electrostatic
field is about 20 for GCT target at $t=60T$, which is about 3.3
times higher than the planar target(6 at $t=44T$). Since the
energetic electrons expand away quickly, the electric fields
decrease quickly after reaching the maximum for both cases. The
electron energy spectrums behind the targets at t=50T for planar
target and t=80T for GCT target are shown in Fig.\ref{fig4}(b), at
both times when the maximum energy of the electrons behind the targets is highest
. We can see that the
electron temperature and density are higher for the GCT target ,
which will result in a higher longitudinal field and eventually
higher proton energy, as shown in Fig.\ref{fig4}(a) and
Fig.\ref{fig3}(a).

Since the TNSA mechanism depends strongly on the charge seperation
field established by the energetic electrons, it is of interest to
investigate the electron number behind the target. Fig.\ref{fig5}(a)
shows the time evolution of electron number behind the planar and
GCT targets. For both cases, the electron numbers initially increase
after the laser irradiates on the targets, and then flatten out. For
GCT target, the electron number is nearly 5 times higher than the
planar target at $t=120T$.
 From Fig.\ref{fig5}(b) we can see the energetic electrons are
almost generated from the gas-plasma ($97\%$ of the total number),
while only a small number of the electrons are from the cone ($3\%$
of the total number). This result indicates that with the GCT target
the efficiency of proton acceleration is determined by the electrons
generated from the gas, which is quite different from the planar
target(the electrons are from the target itself).

\begin{figure}[!ht]
  \includegraphics[width=1.0\columnwidth]{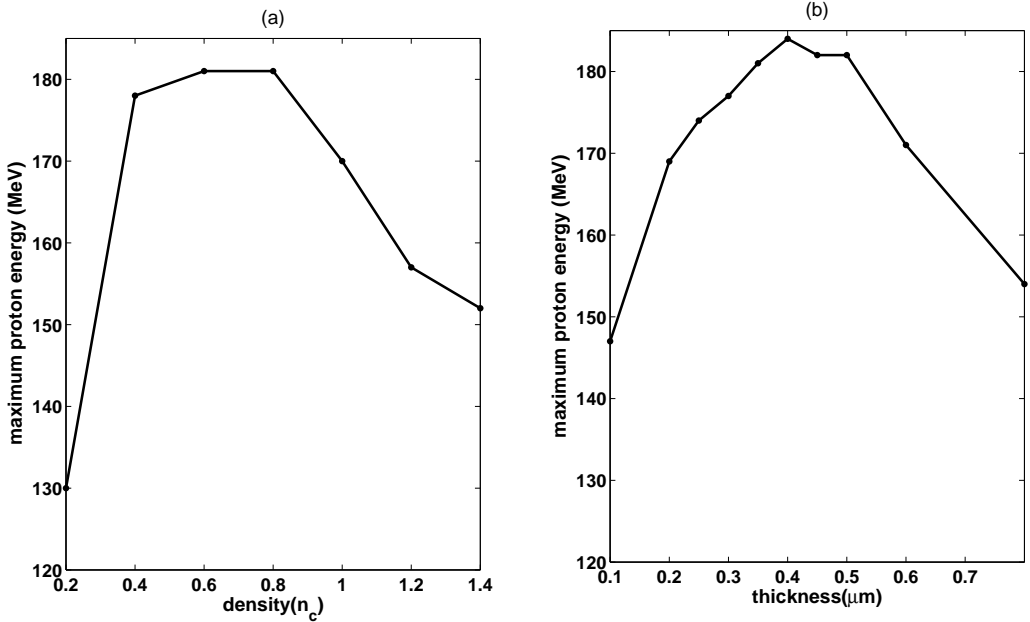}
\caption{(color online).(a) Maximum proton energy for different
gas-plasma density. (b) Maximum proton energy for different foil
thickness. }\label{fig6}
\end{figure}

The effects of the gas-plasma density and the foil thickness of the
GCT target are shown in Fig.\ref{fig6}. It is found that the maximum
proton energy remains almost the same(near 180MeV) while the
gas-plasma density is between $0.4n_c$ and $0.8n_c$, and the foil
thickness is between $0.3\mu m$ and $0.5\mu m$. These simulation
results demonstrate that our acceleration scheme is robust. On
the other hand,  for the gas-plasma density, over-high gas-plasma
density will result in much depletion of laser pulse, while over-low
gas-plasma density will leads fewer enegertic electrons behind the foil,
both will result in a decrease of the maximum proton energy. For the
foil thickness, thin foils proved to be more efficient for ion
acceleration in TNSA by hot-Electron
recirculation\cite{Mackinnon2002}, but over-thin foil will result in
a quick expanding of the electrons, which will also result in a
lower proton energy.

We have also stimulated the interaction of a Linear polarized(LP)
laser pulse with a GCT target at the same intensity, while the other
parameters are the same as in Fig.\ref{fig1}. Our simulations
results verify that similar phenomenon can also be observed with the
LP laser pulse. A quasi-monoenergetic proton beam with peak energy
of 139 MeV and maximum energy of 185 MeV can be generated. This
indicates that our acceleration scheme can be also efficient for the
LP laser pulse.

\section{Conclusion}
In conclusion, proton acceleration from a GCT target is proposed to
enhance the ion energy. A quasi-monoenergetic proton bunch with peak
energy of 130MeV and  maximum energy of 181MeV is achieved by using
the GCT target at laser intensity of $2.6 \times 10^{20}W/cm^2$. It
is nearly three times higher than that from the planar target. This
result is attributed to a stronger electrostatic field behind the
foil, which is formed by the energetic electrons generated and
accelerated by the enhanced laser pulse in the gas-plasma.  The
effects of the gas-plasma density and the foil thickness have been
investigated. The results demonstrate that our acceleration scheme
is robust. Such GCT target may be difficult to make at present,
however, with the rapid advance in nanofabrication technology such a
small conical channel filled with gas-plasma should be
realizable\cite{Ostrikov2005}. Accordingly, the GCT target can
remarkably reduce the cost of a laser driven ion accelerator in the
applications such as cancer therapy.

\begin{acknowledgments}
This work was supported by National Nature Science Foundation of
China (Grant Nos. 10935002,10835003,11025523) and National Basic
Research Program of China (Grant No. 2011CB808104). XQY would like
to thank the support from the Alexander von Humboldt Foundation.
\end{acknowledgments}


%
%
%




\end{document}